\documentclass[aps,prl,preprint,groupedaddress]{revtex4-2}

\usepackage{bm}

\begin{document}


\title{Schr\"{o}dinger representation of quantum mechanics, Berry connection, and superconductivity}


\author{Hiroyasu Koizumi}
\affiliation{Division of Quantum Condensed Matter Physics, Center for Computational Sciences, University of Tsukuba, Tsukuba, Ibaraki 305-8577, Japan}


\date{\today}

\begin{abstract}
The standard quantum mechanical electronic state calculations for molecules and solids uses the Schr\"{o}dinger representation where the momentum conjugate to the coordinate $q_r$ is given by $-i\hbar {\partial \over {\partial q_r}}$.
This formalism contains an extra $U(1)$ phase degree-of-freedom.
We show that it can be regarded as a Berry phase arising from many-electron interaction,
and when it is non-trivial, it gives rise to a current carrying ground state identified as the superconducting ground state. The connection between this superconducting state and the BCS one is presented. 
 \end{abstract}


\maketitle


In the Schr\"{o}dinger's representation explained by Dirac \cite{DiracSec22},
the momenta $p_r$'s conjugate to canonical coordinates $q_r$'s are given by
\begin{eqnarray}
p_r=-i \hbar {{\partial} \over {\partial q_r}}, \quad r=1, \cdots, n
\label{pr1}
\end{eqnarray}
This representation assumes 
the existence of eigenket for the coordinates 
\begin{eqnarray}
|q_1, \cdots, q_n \rangle
\end{eqnarray}

The wave function $\psi(q_1, \cdots, q_n)$ is given using the above ket as
\begin{eqnarray}
\psi(q_1, \cdots, q_n)= \langle q_1, \cdots, q_n| \psi \rangle 
\end{eqnarray}
where $ |\psi \rangle $ is the ket for the physical state.
The wave function must be a single-valued function of the coordinates $q_1,\cdots, q_n$ since the coordinates here are
eigenvalues of the ket $|q_1, \cdots, q_n \rangle$, thus, must be uniquely specified.

From the view point of the Heisenberg formulation of quantum theory \cite{Born-Jordan}, commutation relations 
\begin{eqnarray}
[q_r, q_s]=0, \quad [p_r, p_s]=0, \quad  [q_r, p_s]=i \hbar \delta_{rs}
\end{eqnarray}
are more fundamental than Eq.~(\ref{pr1}). 
The following $p_r$'s are also legitimate 
\begin{eqnarray}
p_r=-i \hbar {{\partial} \over {\partial q_r}}+{{\partial F} \over {\partial q_r}}, \quad r=1, \cdots, n
\label{pr2}
\end{eqnarray}
since they satisfy the same commutation relations.

Dirac claims that we can always use Eq.~(\ref{pr1})\cite{DiracSec22}.
The reason is that the extra term, ${{\partial F} \over {\partial q_r}}$, can be removed
by the change of the wave function
\begin{eqnarray}
\psi(q_1, \cdots, q_n) \rightarrow e^{i \gamma} \psi(q_1, \cdots, q_n) 
\label{gamma}
\end{eqnarray}
where $\gamma$ is related to $F$ by
\begin{eqnarray}
F=\hbar \gamma + \mbox{constant}
\end{eqnarray}
However, this claim has not been carefully examined so far.

In the present work, we examine the phase factor $e^{i \gamma}$ for the ground state.
The standard electronic structure calculation obtains the ground state wave function $e^{i \gamma} \psi(q_1, \cdots, q_n)$ as a whole by employing a finite number of basis functions. In this procedure, $e^{i \gamma}$ is not considered explicitly.

However, there are some cases where the explicit consideration of $e^{i \gamma}$ is necessary due to the fact that the coordinates in the wave function are continuous eigenvalues for continuously parameterized ketvector $|q_1, \cdots, q_n \rangle$. 
Representing a continuous eigenvalue state by a finite number of basis functions may miss something. Such a phenomenon is known in quantum field theory in the context of ``anomaly'' \cite{Fujikawa2004}.
We consider the case where the explicit consideration of $e^{i \gamma}$ is necessary in the following.

Generally, the ground state many-electron wave function for a $N$ electron system can be cast in the following form,
\begin{eqnarray}
\Psi({\bf x}_1, \cdots, {\bf x}_N)=\exp \left(  i \sum_{j=1}^{N} \int_0^{{\bf r}_j} {\bf A}^{\rm MB}_{\Psi}({\bf r}') \cdot d{\bf r}'  \right)
\Psi_0({\bf x}_1, \cdots, {\bf x}_N)
\label{single-valued}
\end{eqnarray}
where ${\bf x}_i$ collectively denotes the coordinate ${\bf r}_i$ and the spin $\sigma_i$ of the $i$th electron, $\Psi_0({\bf x}_1, \cdots, {\bf x}_N)$ is the currentless wave function that is obtained by the energy minimization; the coordinates $q_r, \ (r=1, \cdots, n)$ in Eq.~(\ref{pr1}) correspond to ${\bf r}_i, \ (i=1, \cdots, N)$ with $n=3N$ and from $\Psi$ \cite{koizumi2019,koizumi2020c}.
The factor $\exp \left(  i \sum_{j=1}^{N} \int_0^{{\bf r}_j} {\bf A}^{\rm MB}_{\Psi}({\bf r}') \cdot d{\bf r}'  \right)$ is the one arising from the Berry connection ${\bf A}^{\rm MB}_{\Psi}$ \cite{Berry}, which is defined using $\Psi({\bf x}_1, \cdots, {\bf x}_N)$ by 
 \begin{eqnarray}
{\bf A}^{\rm MB}_{\Psi}({\bf r})&=&-i  \int d\sigma_1 d{\bf x}_{2} \cdots d{\bf x}_{N}
{ {\Psi^{\ast}({\bf r}, \sigma_1, {\bf x}_{2}, \cdots, {\bf x}_{N})} \over {\rho({\bf r})^{{1 \over 2}}}}
 \nabla_{\bf r}
{ {\Psi({\bf r}, \sigma_1, {\bf x}_{2}, \cdots, {\bf x}_{N})} \over {\rho({\bf r})^{{1 \over 2}}}}
\nonumber
\\
&=&{1 \over {\hbar\rho({\bf r})}}\mathfrak{Re}\left(
 \int d\sigma_1  d{\bf x}_{2}  \cdots d{\bf x}_{N}
 \Psi^{\ast}({\bf r}, \sigma_1, \cdots, {\bf x}_{N})
  {\bf p}_{\bf r}
\Psi({\bf r}, \sigma_1, \cdots, {\bf x}_{N}) \right)
\label{Afic}
\end{eqnarray}
where $\rho$ is the electron density obtained 
 \begin{eqnarray}
\rho({\bf r})=\int d\sigma_1 d{\bf x}_{2} \cdots d{\bf x}_{N}\Psi({\bf r},\sigma_1, {\bf x}_{2}, \cdots, {\bf x}_{N})\Psi^{\ast}({\bf r}, \sigma_1, {\bf x}_{2}, \cdots, {\bf x}_{N})
\end{eqnarray}

The comparison of Eqs.~(\ref{gamma}) and (\ref{single-valued}) reads
\begin{eqnarray}
\gamma= \sum_{j=1}^{N} \int_0^{{\bf r}_j} {\bf A}^{\rm MB}_{\Psi}({\bf r}') \cdot d{\bf r}' 
\end{eqnarray}
 This indicates that a non-trivial $e^{i \gamma}$ arises from a non-trivial ${\bf A}^{\rm MB}_{\Psi}$. 

The so-called  ``Bloch's theorem'' states that the ground state is currentless \cite{Bohm1949}.
If this theorem is valid, $\Psi_0$ is the ground state wave function, and ${\bf A}^{\rm MB}_{\Psi}$
is trivial, giving a constant $\exp \left(  i \sum_{j=1}^{N} \int_0^{{\bf r}_j} {\bf A}^{\rm MB}_{\Psi}({\bf r}') \cdot d{\bf r}'  \right)$.
This situation corresponds to a normal ground state. 
However, if ${\bf A}^{\rm MB}_{\Psi}$ is non-trivial, a current carrying ground state may arise.
We may identify such a ground state as a superconducting one. 

Since ${\bf A}_{\Psi}^{\rm MB}$ is a self-referencing quantity obtained from the wave function $\Psi$ itself as seen in Eq.~(\ref{Afic}),
we employ general requirements to derive it from $\Psi_0$ obtained by the usual procedure.

The requirements are
\begin{enumerate}

\item The single-valuedness of $\Psi({\bf x}_1, \cdots, {\bf x}_N)$ as a function of ${\bf r}_1, \cdots, {\bf r}_N$.

\item The local charge conservation.

\end{enumerate}

The first requirement is necessary since we use Eq.~(\ref{pr1}) which assumes the uniqueness of $q_r$'s.

The second requirement is a usual requirement for electron systems. Actually, it is equivalent to the energy minimization requirement with respect to the variation of $\gamma$.

Our previous investigations show that current carrying ground states can be obtained using the above requirements \cite{koizumi2020c,Koizumi2021c}.

Now, we treat $\gamma$ as a field. For this purpose, it is more convenient to use $\chi$  
given by
\begin{eqnarray}
{ {\chi({\bf r})}}= - 2\int^{{\bf r}}_0 {\bf A}_{\Psi}^{\rm MB}({\bf r}') \cdot d{\bf r}' 
\end{eqnarray}

The presence of $\chi$ makes the total energy as a functional of $\chi$, which we denote as $E[\chi]$.
This $\chi$ enters only in the kinetic energy part through $\nabla \chi$ by the replacement 
\begin{eqnarray}
-i\hbar \nabla \rightarrow -i \hbar \nabla -{ \hbar \over 2} \nabla \chi
\end{eqnarray}
This indicates that $\nabla \chi$ gives rise to a vector potential for a ``fictitious magnetic field''
\begin{eqnarray}
{\bf A}^{\rm fic}={{c\hbar} \over {2(-e)}} \nabla \chi
\end{eqnarray}
where $-e$ is the electron charge.

Then, using a general formula for the current from the energy functional, 
the current density is 
\begin{eqnarray}
{\bf j}=-c {{ \delta E} \over {\delta {\bf A}^{\rm fic}}}=-
{ {2e}  \over \hbar} {{ \delta E} \over {\delta \nabla \chi}}
\end{eqnarray}

The fact that $E[\chi]$ only depends on $\nabla \chi$ yields the following relation from the minimization requirement of the total energy with respect to $\chi$
\begin{eqnarray}
0={{ \delta E} \over {\delta  \chi}}=-\nabla \cdot {{ \delta E} \over {\delta \nabla \chi}}
={ \hbar \over {2e} }\nabla\cdot {\bf j}
\end{eqnarray}
which proves the energy minimization is equivalent to the local charge conservation as mentioned before.

Let us consider the quantization of the field $\chi$.
The canonical conjugate momentum of $\chi$, denoted by $\pi_{\chi}$, is obtained from the Lagrangian
\begin{eqnarray}
L&=&\langle \Psi | i\hbar {\partial \over {\partial t}} -H |\Psi \rangle
\nonumber
\\
&=&\int d^3 r \hbar { \dot{\chi} \over 2}\rho+ i\hbar \langle \Psi_0 |{\partial \over {\partial t}}|\Psi_0 \rangle-\langle \Psi | H |\Psi \rangle
\end{eqnarray}
as
\begin{eqnarray}
\pi_{\chi}={{\delta L} \over {\delta \dot{\chi} }}={ \hbar \over 2} \rho
\end{eqnarray}

The canonical quantization condition reads
\begin{eqnarray}
[\chi({\bf r}),\pi_{\chi} ({\bf r}')]=i\hbar \delta({\bf r}-{\bf r}')
\end{eqnarray}
or
\begin{eqnarray}
[\chi({\bf r}),\rho({\bf r}')]=2i\delta({\bf r}-{\bf r}')
\end{eqnarray}

Then, we can construct the following boson field operators 
\begin{eqnarray}
\psi_{\chi}^{\dagger}({\bf r})=\sqrt{\rho({\bf r})}e^{{ i \over 2}
\chi({\bf r})}, \quad \psi_{\chi}({\bf r})=e^{-{ i \over 2}\chi({\bf r})}\sqrt{\rho({\bf r})}
\end{eqnarray}
that satisfy the commutation relation
$
[ \psi_{\chi}({\bf r}), \psi_{\chi}^{\dagger}({\bf r}')]=\delta({\bf r}-{\bf r}')
$.

Integrating $\psi_{\chi}^{\dagger}({\bf r})$ and  $\psi_{\chi}({\bf r})$ over the space, we obtain boson operators $B^{\dagger}_{\chi}$ and $B_{\chi}$
given by
\begin{eqnarray}
B^{\dagger}_{\chi} =\int d^3r  \psi_{\chi}^{\dagger}({\bf r}),
\quad B_{\chi} =\int d^3r \psi_{\chi}({\bf r})
\end{eqnarray}
They satisfy the following commutation relation
\begin{eqnarray}
[ B_{\chi}, B_{\chi}^{\dagger}]=1
\end{eqnarray}

We define the number operator 
\begin{eqnarray}
\hat{N}_{\chi}=B_{\chi}^{\dagger} B_{\chi}
\end{eqnarray}
and introduce the eigenket $|{N}_{\chi} \rangle$,
\begin{eqnarray}
\hat{N}_{\chi}|{N}_{\chi} \rangle
={N}_{\chi}|{N}_{\chi} \rangle
\end{eqnarray}
The number ${N}_{\chi}$ can be considered as the number of electrons participating in the collective mode described by $\chi$. 

A phase operator $\hat{X}$ that is conjugate to $\hat{N}_{\chi}$ is defined 
 through the relations
 \begin{eqnarray}
B_{\chi}^{\dagger} =\sqrt{\hat{N}_{\chi}}e^{i{1 \over 2}\hat{X}}, \quad B_{\chi} =e^{-i{1 \over 2}\hat{X}}
\sqrt{\hat{N}_{\chi}}
\end{eqnarray}

The phase and number operators satisfy the following commutation relations
\begin{eqnarray}
[\hat{N}_{\chi}, e^{\pm i {1 \over 2}\hat{X}}] =\pm e^{\pm i {1 \over 2}\hat{X}}
\end{eqnarray}

We can derive the following relations
\begin{eqnarray}
e^{\pm i {1 \over 2}\hat{X}} |{N}_{\chi} \rangle \propto |{N}_{\chi} \pm 1\rangle
\end{eqnarray}
Thus, 
$e^{ -i{1 \over 2} \hat{X}}$ is the number changing operator that decreases the
number of electrons participating in the collective mode by one, $e^{i{1 \over 2}\hat{X}}$ increases by one.

We now reformulate the standard theory of superconductivity, the BCS theory, using the number changing operators $e^{\pm i {1 \over 2}\hat{X}}$ According to this theory,
the origin of superconductivity is the energy gap formation due to electron-pairing \cite{BCS1957}. The rigidity of the wave function against external perturbations envisaged by London \cite{London1950} is realized by this energy gap, and one of the hallmarks of superconductivity, the exclusion of a
magnetic field from the superconductor (the Meissner effect), is explained by this rigidity.

The BCS used the following variational state vector, 
\begin{eqnarray}
|{\rm BCS}\rangle =
\prod_{\bf k}(u_{\bf k}+e^{i\theta}v_{\bf k}c^{\dagger}_{{\bf k} \uparrow} c^{\dagger}_{-{\bf k} \downarrow} )|{\rm vac} \rangle
\label{theta}
\end{eqnarray}
to take into account the electron pairing effect,
 where $|{\rm vac} \rangle$ is the vacuum state, $c^{\dagger}_{{\bf k} \sigma}$ is the creation operator for the conduction electron of effective mass $m^{\ast}$ with the wave vector ${\bf k}$ and spin $\sigma$, and $u_{\bf k}$ and  $v_{\bf k}$ are variational parameters. The obtained energy gap explains many experimental results, and a method to calculate the superconducting transition temperatures is provided as the method to calculate the energy gap formation temperature \cite{BCS1957}. 

A salient feature of the BCS state vector in Eq.~(\ref{theta}) is that it dose not satisfy the conservation of the particle number. This is odd since superconductivity occurs in an isolated superconductor where the number of electrons is fixed \cite{LeggettBook}.
However, this non-conservation property is a crucial ingredient; it makes the phase factor $e^{i\theta}$ physically meaningful. This breaks the global $U(1)$ gauge symmetry, and this gauge symmetry breaking is needed to explain the Meissner effect in superconductors \cite{Anderson1958a,Anderson1958b,Nambu1960}.

Using the number changing operator $e^{ i\hat{X}}$, a state vector similar to the one in Eq.~(\ref{theta}) is constructed 
\begin{eqnarray}
|{\rm Gnd}\rangle =
\prod_{\bf k}(u_{\bf k}+v_{\bf k} c^{\dagger}_{{\bf k} \uparrow} c^{\dagger}_{-{\bf k} \downarrow}e^{-i \hat{X}} )|{\rm Cnd} \rangle
\label{chi}
\end{eqnarray}
 where the state vector $|{\rm Cnd} \rangle$ corresponds to the state given by the wave function $\Psi$. This state corresponds to the state in which all the electrons participate in the collective mode
 \begin{eqnarray}
|{\rm Cnd} \rangle = |{N}_{\chi}=N \rangle
\end{eqnarray}

The operator
$c^{\dagger}_{{\bf k} \uparrow} c^{\dagger}_{-{\bf k} \downarrow}e^{i \hat{X}}$
 acting on $|{\rm Cnd} \rangle$ decreases the number of electrons participating in the collective mode by two, and increases
 the number of  electrons in single-particle motion. Thus, the particle number is fixed in Eq.~(\ref{chi}).
 
Let us consider an example using Eq.~(\ref{chi}).
It is a two dimensional system with the following single particle Hamiltonian
\begin{eqnarray}
h=-{{\hbar^2} \over {2 m_e}}(\partial_x^2 +\partial_y^2)+U(r)
\end{eqnarray}
 For simplicity we assume that the potential $U$ depends only on $r$ with $x=r \cos \phi$ and $y=r \sin \phi$.

The coordinate part of the wave function is a product of an angular function and a radial function given by
\begin{eqnarray}
\varphi_{nm}(r, \phi)={1 \over \sqrt{2 \pi}} e^{ i m \phi} R_{n |m|}(r)
\end{eqnarray}
where $m$ is an integer, $n$ is a natural number that denotes the number of nodes of the radial wave function, $R_{n |m|}(r)$; $\varphi_{nm}(r, \phi)$ is the eigenfunction of $h$ with energy $E_{n m}$, 
\begin{eqnarray}
 h\varphi_{nm}(r, \phi)=E_{n m}\varphi_{nm}(r, \phi)
\end{eqnarray}

Usually, the wave functions
$\psi_{n m \uparrow}= \varphi_{n m}(r, \phi)|\uparrow \rangle$ and  $\psi_{n m \downarrow}=\varphi_{n m}(r, \phi)|\downarrow \rangle$
are used by adopting the coordinate independent spin functions $|\uparrow \rangle$
and $|\downarrow \rangle$.
However, we consider the following spin functions
\begin{eqnarray}
|\Sigma_a  \rangle&=&{ 1 \over \sqrt{2}}(e^{-{i \over 2} f(\phi)} \sin \zeta_0 |\uparrow \rangle 
+e^{{i \over 2}f(\phi)} \cos \zeta_0  |\downarrow \rangle ),
\nonumber
\\
|\Sigma_b \rangle&=&{ 1 \over \sqrt{2}}(-e^{-{i \over 2}f(\phi)} \sin \zeta_0  |\uparrow \rangle 
+e^{{i \over 2}f(\phi)}  \cos \zeta_0|\downarrow \rangle )
\label{spin}
\end{eqnarray}
where $f(\phi)$ is a function of $\phi$, and $\zeta_0$ is a constant,
and use the following wave functions,
\begin{eqnarray}
\tilde{\psi}_{n m a}= \varphi_{n m}(r, \phi)|\Sigma_a  \rangle, \quad \tilde{\psi}_{n m b}=\varphi_{n m}(r, \phi)|\Sigma_b \rangle
\label{tilde}
\end{eqnarray}

Expectation values of the components of spin ${\bf s}=(s_x, s_y, s_z)$ for $|\Sigma_a \rangle$ are given by
\begin{eqnarray}
\langle \Sigma_a |s_x|\Sigma_a \rangle &=& { \hbar \over 2}\cos f(\phi) \sin \zeta_0,
\nonumber
\\
\langle \Sigma_a |s_y|\Sigma_a \rangle &=&{ \hbar \over 2}\sin f(\phi) \sin \zeta_0,
\nonumber
\\
\langle \Sigma_a |s_z|\Sigma_a \rangle &=&{ \hbar \over 2} \cos \zeta_0
\end{eqnarray}
and those for $|\Sigma_b \rangle$ are 
$
\langle \Sigma_b |{\bf s}|\Sigma_b \rangle=-\langle \Sigma_a |{\bf s}|\Sigma_a\rangle
$.

We consider the case where spin-twisting occurs around the $z$-axis. For simplicity, we consider the following case
\begin{eqnarray}
f(\phi)=\phi
\end{eqnarray}
Then, the spin functions in Eq.~(\ref{spin}) become multi-valued as follows
\begin{eqnarray}
\phi \rightarrow \phi + 2\pi ; \quad \Sigma_a \rightarrow -\Sigma_a, \quad \Sigma_b \rightarrow -\Sigma_b
\end{eqnarray}
Note that $\phi$ and $\phi + 2\pi$ describe the same coordinate, however, two values of $\Sigma_a$ and $\Sigma_b$ arise.
The wave functions in Eq.~(\ref{tilde}) also show the same multi-valuedness.

Now we take into account the phase factor $e^{i \gamma}$. The wave functions including the phase factor $e^{i \gamma}$ are given by
\begin{eqnarray}
{\psi}_{n m a}=\tilde{\psi}_{n m a}e^{-i { \chi \over 2}}, \quad
{\psi}_{n m b}=\tilde{\psi}_{n m b}e^{-i { \chi \over 2}} 
\end{eqnarray}
The above new wave functions are single-valued if the phase factor $e^{i { \chi \over 2}}$ compensates the
sign change of $\tilde{\psi}_{n m a}$ and $\tilde{\psi}_{n m b}$.
Since the multi-valuedness of $\tilde{\psi}_{n m a}$ and $\tilde{\psi}_{n m b}$ arise from their $\phi$ dependence, we may treat $\chi$ as a function of $\phi$.

The total energy $E[\chi]$ depends on $\chi$ through ${{d \chi} \over {d \phi}}$. 
Thus, the condition for an optimal $\chi$ that minimizes the total energy yields
$
\chi=A \phi +B 
$,
where $A$ and $B$ are constants. We can put $B=0$ since it merely gives rise to a constant phase factor on the wave function.
The constant $A$ must be so chosen that $e^{{i \over 2}(A\pm1)\phi}$ is a single-valued function of the coordinate. 
Then, $A$ ia an odd integer, and the minimal energy one will be $A=1$ or $A=-1$.

The total energy becomes the sum of the kinetic energy from $e^{i \gamma}$ and the total energy from $\psi(q_1, \cdots, q_n)$ \cite{Bohm1949} due to the fact that the current is zero for $\psi(q_1, \cdots, q_n)$.
Thus, we have
\begin{eqnarray}
E[\chi]=\int dr d \phi  {{\rho(r)\hbar^2}\over {2 m_e}}
\left({ 1 \over 2}{{d \chi}  \over {d \phi}} \right)^2
+\sum_{\tilde{E}_{n m} \leq 0}2 \tilde{E}_{n m}
\end{eqnarray}
where $\tilde{E}_{n m}={E}_{n m}-E_F$ and $E_F$ is the Fermi energy. The factor $2$ in the second term appears due to the fact that both $\psi_{n m a}$ and $\psi_{n m b}$ are occupied.
This current carrying state is energetically higher than the currentless state  \cite{Bohm1949}. 

Let us introduce the pairing interaction given by
\begin{eqnarray}
 H_{\rm pair}&=&\sum_{n, m, n',m'} V_{n m; n' m'} c^{\dagger}_{n m \uparrow} c^{\dagger}_{n -m \downarrow}
c_{n' -m' \downarrow} c_{n' m' \uparrow} 
\nonumber
\\
&=&\sum_{n, m, n',m'} V_{n m; n' m'} c^{\dagger}_{n m \uparrow} c^{\dagger}_{n -m \downarrow}  e^{-i\hat{X}} e^{i\hat{X}}
c_{n' -m' \downarrow} c_{n' m' \uparrow} 
\end{eqnarray}
where $c^{\dagger}_{n m \sigma}$ and $c_{n m \sigma}$ are creation and annihilation operators for
$ \varphi_{n m}(r, \phi)|\sigma\rangle$,
and the identity
$
1= e^{-i\hat{X}} e^{i\hat{X}}
$ is used.

The total Hamiltonian is the sum of $H_{\rm pair}$ and the single-particle Hamiltonian given by
\begin{eqnarray}
 H_0=\sum_{n, m} \tilde{E}_{n m}( c^{\dagger}_{n m \uparrow} c_{n m \uparrow} +c^{\dagger}_{n m \downarrow} c_{n m \downarrow})
\end{eqnarray}

We employ a BCS type variational ground state 
\begin{eqnarray}
|{\rm Gnd}\rangle =\prod_{n,m}(u_{nm}+v_{nm}c^{\dagger}_{n m \uparrow} c^{\dagger}_{n -m \downarrow}
 e^{-i\hat{X}})|{\rm Cnd} \rangle
 \label{eqGnd}
\end{eqnarray}
where 
$u_{nm}$ and $v_{nm}$ are variational parameters that satisfy
$u_{nm}^2+v_{nm}^2=1$.

The pairing energy gap is defined as
\begin{eqnarray}
 \Delta_{n m}&=&-\sum_{n, m, n',m'} V_{n m; n' m'} \langle {\rm Gnd}| e^{i\hat{X}}c_{n' -m' \downarrow} c_{n' m' \uparrow} |{\rm Gnd}\rangle
 \nonumber
 \\
 &=&-\sum_{n, m, n',m'} V_{n m; n' m'} u_{n' m'}v_{n' m'}
\end{eqnarray}
where 
$e^{i\hat{X}}c_{n' -m' \downarrow}c_{n' m' \uparrow}$ annihilates two electrons in the single-particle mode and creates two electrons in 
the collective mode with conserving the particle number.

As in the BCS theory, we assume $V_{n m; n' m'}=-V$ if $|\tilde{E}_{n m}|, |\tilde{E}_{n' m'}| \leq \hbar \omega_c$ and zero otherwise, where $\hbar\omega_c$ is a cut-off energy \cite{BCS1957}.
Then, we obtain
\begin{eqnarray}
u_{n m}^2={ 1 \over 2} \left(1 +{{\tilde{E}_{n m}} \over {\sqrt{ \tilde{E}_{n m}^2+ \Delta^2}}} \right), \quad
v_{n m}^2={ 1 \over 2} \left(1 -{{\tilde{E}_{n m}} \over {\sqrt{ \tilde{E}_{n m}^2+ \Delta^2}}} \right)
\end{eqnarray}
where the pairing energy gap is given by 
\begin{eqnarray}
\Delta\approx 2 \hbar \omega_c e^{-{ 1 \over {N(0)V}}}
\end{eqnarray}
with $N(0)$ being the density of states at the Fermi energy.

The total energy becomes
\begin{eqnarray}
E_{\rm tot}&=&\int r dr d \phi  {{\rho(r)\hbar^2}\over {2 m_e}}
\left({ 1 \over 2}{{d \chi}  \over {d \phi}} \right)^2+2\sum_{m n} \tilde{E}_{n m} v_{n m}^2-{{\Delta^2} \over V}
\nonumber
\\
&=&\int r dr d \phi  {{\rho(r)\hbar^2}\over {2 m_e}}
\left({ 1 \over 2}{{d \chi}  \over {d \phi}} \right)^2-{1 \over 2}N(0)V\Delta^2
\end{eqnarray}
where the number of electrons in the collective mode is calculated as
\begin{eqnarray}
\int  rdr d \phi  \rho(r) &=& \sum_{ \tilde{E}_{n m}\leq 0}u^2_{n m}
\nonumber
\\
&=&N(0)\left(\hbar \omega_c +\sqrt{\Delta^2}-\sqrt{ \hbar^2 \omega_c^2+ \Delta^2} \right)
\nonumber
\\
&\approx& N(0)\left(\Delta-{ 1 \over {2 \hbar \omega_c}}\Delta^2 +{1 \over {8\hbar^3 \omega_c^3}}\Delta^4 \right)
\label{nonzero}
\end{eqnarray}

If the energy gap formation makes the current carrying state lower in energy than the currentless state, the superconducting state is realized. This example only consider one centers of spin-twisting; in reality,
multi-spin-twisting centers are more energetically favorable. 
If we consider a more general setting by including the potential energy from the underlying ion lattice and effective field from other electrons, the spin-twisting itinerant motion occurs as the circular motion around a section of the Fermi surface of the metal \cite{koizumi2020}. This will correspond to a system with multi-spin-twisting-centers in the coordinate space.

When a magnetic field exists, the vector potential from magnetic field ${\bf A}^{\rm em}$ appears in addition. Then, the kinetic energy of the collective mode is given by
\begin{eqnarray}
E_{\chi}=\int d^3 r {{\hbar^2 \rho({\bf r})} \over {2 m_e}}
\left({ 1 \over 2}\nabla \chi-{e \over {c\hbar}}{\bf A}^{\rm em}
\right)^2
\end{eqnarray}

Then, the supercurrent density is given by
\begin{eqnarray}
{\bf j}=-c{{\partial E_{\chi}} \over {\partial {\bf A}^{\rm em}}}
=-{{e^2 \rho({\bf r})} \over {m_e c}}
\left({\bf A}^{\rm em}-{ {c\hbar} \over {2e}}\nabla \chi
\right)
\end{eqnarray}
This is a diamagnetic current explains the Meissner effect. Note that the ambiguity in the gauge of ${\bf A}^{\rm em}$ is absorbed during the optimization of $\nabla \chi$, thus, the current is gauge invariant.
The period $2\pi$ of the angular variable $\chi$ yields the flux quantum ${ {ch} \over {2e}}$. 

The velocity field associated with the above supercurrent is 
\begin{eqnarray}
{\bf v}_s
={{e} \over {m_e c}}
\left({\bf A}^{\rm em}-{ {c\hbar} \over {2e}}\nabla \chi
\right)
\label{LondonM}
\end{eqnarray}
The velocity field generated inside the superconductor by rotating it with an angular velocity ${\bm \omega}$ is given by
$
{\bf v}_{\rm rot}
={\bm \omega} \times {\bf r}
$.
Since supercurrent electrons move with the body to shield the electric field from the ion core, the condition ${\bf v}_s={\bf v}_{\rm rot}$ is satisfied. Substituting this in Eq.~(\ref{LondonM}) yields the magnetic field 
\begin{eqnarray}
{\bf B}^{\rm em}={{2m_e c} \over e} {\bm \omega}
\label{LondonF}
\end{eqnarray}
inside the superconductor. This is known as the London moment phenomenon.
 The above formula has the free electron mass in accordance with the experimental results.
Note that the effective mass $m^{\ast}$ is used for the supercurrent carrier in the BCS theory, thus, the mass in Eq.~(\ref{LondonF}) becomes $m^{\ast}$ instead of $m_e$ in disagreement with the experiment. This point
has been discussed by some researchers \cite{Hirsch2013b,koizumi2021}. 

The present work indicates the superconductivity is a phenomenon where the non-trivial ${\bf A}^{\rm MB}_{\Psi}$ appears.
The electron pairing stabilizes it, and the BCS theory takes into account this stabilization effect using the variational wave function that breaks  the global $U(1)$ gauge symmetry. However, the global $U(1)$ gauge symmetry breaking is not necessary for
superconductivity but the non-trivia $e^{i \gamma}$ is.

%


\end{document}